\documentclass[12pt,preprint]{emulateapj}
\usepackage{graphicx}
\usepackage{times}
\usepackage{apjfonts}
\usepackage{isotope}
\shorttitle{NIR Gas cells for Precision Velocity}
\shortauthors{Mahadevan \& Ge}
\begin{document}
\title{On The Use of Absorption Cells as a Wavelength Reference for Precision Radial Velocity Measurements in the Near Infrared}
\author{Suvrath Mahadevan\altaffilmark{1} \& Jian Ge\altaffilmark{1}}
%  \newauthor % starts a new line in the
%             % author environment
%  A.~Richardson,$^1$
%  P.~Smith,$^2$\thanks{Production Editor.}
%  N. Thompson$^2$\footnotemark[2]
%  and G. Hutton$^2$\footnotemark[2] \\
%\\
%  $^1$Department of Astronomy, University of Florida, 211 Bryant Space Science Center, %Gainesville, Florida}
\altaffiltext{1}{Department of Astronomy, University of Florida, 211
  Bryant Space Science Center, Gainesville, FL 32611-2055}
%\author[Mahadevan et al.]{Suvrath Mahadevan$^1$\footnotemark[1] et al.
%\\
%$^1$Department of Astronomy, University of Florida\\}
%\affil{Astronomy Department,  University of Florida, 211 Bryant
%Space Science Center P.O. Box 112055 Gainesville, FL 32611-2055}
%\email{suvrath@astro.ufl.edu}

%\altaffiltext{2}{Society of Fellows, Harvard University.}
%\altaffiltext{3}{present address: Center for Astrophysics,
%    60 Garden Street, Cambridge, MA 02138}
%\altaffiltext{4}{Visiting Programmer, Space Telescope Science Institute}
%\altaffiltext{5}{Patron, Alonso's Bar and Grill}
%\date{Released 2002 Xxxxx XX}

%\pagerange{\pageref{firstpage}--\pageref{lastpage}} \pubyear{2008}

%\def\LaTeX{L\kern-.36em\raise.3ex\hbox{a}\kern-.15em
%    T\kern-.1667em\lower.7ex\hbox{E}\kern-.125emX}

%\label{firstpage}
%\maketitle
\begin{abstract}
Considerable interest is now focused on the detection of terrestrial mass planets around M dwarfs, and radial velocity surveys with high-resolution spectrographs in the near infrared are expected to be  able to discover such planets.
We explore the possibility of using commercially available molecular absorption gas cells as a wavelength reference standard for high-resolution fiber-fed spectrographs in the near-infrared. We consider the relative merits and disadvantages of using such cells compared to Thorium-Argon emission lamps and conclude that in the astronomical H band they are a viable method of simultaneous calibration, yielding an acceptable wavelength calibration error for most applications. Four well-characterized and commercially available  standard gas cells of H\isotope[13]{C}\isotope[14]{N}, \isotope[12]{C}$_2$H$_2$, \isotope[12]{CO}, and \isotope[13]{CO} can together span over 120nm of the H band, making them suitable for use in astronomical spectrographs. The use of isotopologues of these molecules can increase line densities and wavelength coverage, extending their application to different wavelength regions.
\end{abstract}


%\begin{keywords}
%techniques: radial velocities --- techniques:spectroscopic ---  instrumentation: %interferometers ---instrumentation: spectrographs --- stars: kinematics --- %methods:data analysis
%\end{keywords}
\section{INTRODUCTION}
With the discovery of more than 300 extrasolar planets, considerable interest is now focused on finding and characterizing terrestrial mass planets in habitable zones around their host stars. Such planets are extremely difficult to detect around F, G and K stars, requiring either extremely high radial velocity precision ($< 1$ m s$^{-1}$) or space based photometry to detect a transit. Radial velocity and transit efforts are now beginning to focus on M dwarfs, the most numerous stars, where the lower luminosity shifts the habitable zone much closer to the star. Terrestrial mass planets in such regions are detectable with the current radial velocity precision obtained with high resolution echelle spectrographs \citep{Butler96, Pepe04}. A number of the brighter M dwarfs are being surveyed in the optical with existing high precision radial velocity instruments. Neptune and super-Earth mass planets have been discovered around a number of such objects. Examples are GJ436 (Butler et al. 2004), GL581 (Udry et al. 2007), and GL176 (Forveille et al. 2008) and they suggest that such planets may be rather common around M stars. The habitability of terrestrial planets around M stars has been explored by \citet{Tarter07} and \citet{Scalo07}.

However, most of these stars are intrinsically faint in the optical, emitting most of their flux in the 1-1.8 $\mu m$ wavelength region, the near infrared (NIR) J (1.1-1.4 $\mu m$) and H (1.45-1.8 $\mu m$) bands. Stellar activity, coupled with the relative faintness, can make detections in the optical difficult. For example, Endl et al. (2008) claim the detection of a 24 Earth-mass planet around Gl 176 using 28 radial velocities obtained with the High Resolution Spectrograph (HRS) on the Hobby-Eberley Telescope, while Forveille et al. (2008), with 58 higher precision measurements obtained with the HARPS (High Accuracy Radial velocity Planet Searcher) instrument, find evidence for stellar activity as well as a different orbital period and a much smaller minimum mass. Similar observation in the NIR may require less telescope time and are less prone to systematics because the activity induced radial velocity jitter is expected to be smaller in the NIR than in the optical. As an example, Setiawan et al. (2008) announced the discovery of a planet around the young active star TW Hya, but Huelamo et al. (2008) see no such variability on precision radial velocity data obtained with the CRIRES (Cryogenic Infrared Echelle Spectrograph) instrument in the infra-red (Seifhart \& Kaufl 2008) and they attribute the velocity variability in the optical to star spots. Rucinski et al. (2008) also observe photometric periodicity at the proposed planet's period in one season of observations with the MOST (Microvariability $\&$ Oscillations of Stars) satellite, and the variations are absent in another season- suggesting that the cause of the radial velocity is activity-induced. Prato et al. (2008) have discussed radial velocity observations of young stars in the optical and NIR and conclude that observations in the near-infrared are essential to discriminate between activity and the presence of planets. NIR spectroscopy can explore new regimes of planet formation as well as complement existing observational programs in the optical that target young or active stars.

In this article, we explore the calibration challenges faced by a high-resolution spectrograph operating in the NIR and suggest that easily available absorption gas cells can be used for wavelength calibration purposes for any such instruments that are currently being designed. Our discussion focuses on using a simultaneous wavelength calibrator along a separate optical fiber \citep{Baranne96} to track and calibrate out instrument drift, similar to the Th-Ar calibration technique in the optical. We briefly highlight important aspects on such instruments in \S2. In \S3 and \S4 we discuss currently used emission wavelength sources and their relative merits and disadvantages. In \S5 we explore the use of commercially available absorption cells for precision wavelength calibration.
\section{Fiber-Fed High Resolution Spectrographs in the NIR}
The use of optical fibers allows a spectrograph to be placed in a stable environmentally controlled enclosure. The intrinsic radial scrambling properties of an optical fiber can, with the use of a double scrambler \citep{HR92}, lead to a very stable illumination profile on the spectrograph slit. Commercially available optical fibers provide high transmission in the J and H NIR bands. Fiber-feeds allow a second calibration fiber to be used to simultaneously track the instrument drift during an object exposure. If the science and calibration fiber images are relatively close on the focal plane, then the measured drift of the calibration spectrum is a very good estimate of the drift estimated on the science spectrum. The major requirements on the calibration source are that it be stable enough to achieve the required precision, have a number of lines with known wavelengths, and be bright enough that sufficient S/N is obtained for it to not compromise the possible radial velocity accuracy on bright sources. Design studies for such instruments in the NIR have been conducted, including one for a proposed Precision Radial Velocity Spectrograph \citep{Rayner07} for the Gemini telescope. On the fiber-fed bench top PRVS Pathfinder instrument, \citet{Ramsey08} have demonstrated a short term radial velocity precision of 7-10 m s$^{-1}$ in the NIR using integrated sunlight with Thorium-Argon as a wavelength reference. The use of silicon immersion gratings is also being explored (eg. Ge at al. 2006, Jaffe et al. 2006) to facilitate compact spectrographs capable of providing spectral resolutions of R=50-100k in the NIR.
\section{Thorium-Argon Emission Lamps: }
\subsection{History \& Advantages} \citet{Kerber07} provide a summary of the development and use of Thorium--Argon (Th-Ar) hollow cathode emission lamps. The only naturally occurring Thorium (Th) isotope, \isotope[232]{Th}, has zero nuclear spin, leading to sharp symmetric emission lines, even at very high resolutions. The monatomic Argon gas also has a number of bright lines. Th-Ar emission lines span the UV-NIR regions, making such a lamp a very useful and convenient source of wavelength calibration. Most wavelength calibration applications using this lamp are tied to the \citet{PE83} (PE83 hereafter) measurements of the Thorium lines in the 280-1100nm region. Their quoted measurement precision, $\sim$ 0.001 cm$^{-1}$ - 0.005 cm$^{-1}$, corresponds to an intrinsic velocity uncertainty of 15-80 ms$^{-1}$ per emission line (at 550nm). For Argon (Ar) lines, the wavelength measurements typically used are adopted from \citet{Norlen73}. More recent measurements of Ar lines have been taken by Whaling et al. (1995, 2002), who extend measurements to the mid-IR. Argon is not a heavy element like Thorium and the central wavelengths of the Argon emission lines are susceptible to pressure shifts, making them sensitive to environmental conditions in the lamp. The lines may be unstable by many tens of ms$^{-1}$ \citep{LP07}, and for the highest possible precision ($\sim 1-3$  ms$^{-1}$), Ar lines should generally be avoided. At intermediate spectral resolutions, a further problem is line blends, because the intensity ratios of Th and Ar lines is a function of the current supplied \citep{Kerber07}. Very stable spectrographs offer the opportunity of decreasing the internal errors associated with the original Thorium measurements and PE83s use of the Fourier Transform Spectrograph (FTS) at Kitt-Peak National Observatory (KPNO). Stable instruments allow one to acquire many Th-Ar spectra and co-add them, thereby increasing the S/N and reducing the photon noise uncertainties on the line-centers of known lines, as well as enabling wavelength estimation of lines not initially present in the original PE83 atlas.
Such a technique has been applied by \citet{LP07} with the vacuum enclosed HARPS high-resolution echelle spectrograph, resulting in a line-list with improved wavelength precision (though inheriting the zero-point of the original KPNO atlas). This improved line-list has certainly helped HARPS achieve radial velocity precisions of $<$1 ms$^{-1}$ and discover super-Earths \citep{Mayor08}.

\subsection{Calibration in the NIR} In the NIR J and H bands Th lines are few and relatively faint, making it difficult to find accurate dispersion solutions in the J and H band without substantial integration time. The \citet{Hinkle01} atlas contains FTS and grating spectrograph wavelength measurements of ~ 500 lines in the 1-2.5 $\mu m $ range and Engleman, Hinkle \& Wallace (2003) derive FTS wavelengths for a much larger sample using a cooled Th-Ar cell operated at a high current. \citet{Kerber08} also measured wavelength for Th-Ar lines for multiple cells with an FTS for use in calibration of the CRIRES instrument model. Although these atlases help to identify lines when they are detected, the Th lines in commercially-available lamps are still very low in intensity. The high contrast between the Th and Ar lines leads to the bright Ar lines dominating the spectra and these can be a source of significant scattered light. For example, the SOPHIE instrument (Perruchot et al. 2008) requires the use of a ~700 nm shortpass filter to block scattered light from bright Ar lines in the infrared. For high precision radial velocity studies, the situation is further exacerbated by the need to acquire a high enough signal level on sufficient lines to achieve a dispersion solution accurate enough to be able to measure radial velocities at a high precision. Another minor concern is that simultaneous calibration with Th-Ar also makes scattered light subtraction of echelle frames difficult. The need to have an acceptable signal on most lines leads to the strongest lines saturating and bleeding into the adjacent stellar spectrum, making reduction more complex. The relative lack of sufficient Th lines in the NIR has also been noted by \citet{Ramsey08} in their laboratory tests with PRVS Pathfinder and these authors propose combining light from multiple lamps (most notable Uranium-Ar) with Th-Ar to increase the line density available for wavelength calibration.
\section{Broadband Frequency Combs}
The calibration advantages of using a laser frequency comb with high resolution spectrographs have been discussed by \citet{Murphy07} and \citet{Braje08}.
Though these have the desired frequency coverage, the intrinsically small frequency mode spacing (250MHz-- 1GHz) makes their direct use unsuitable for astronomical spectrographs since the individual lines would blur together. The use of a Fabry Perot filter to select only well-spaced emission lines has been demonstrated by \citet{Li08} for optical wavelengths and in parts of the NIR H band (1530-1600nm) by Steinmetz et al. (2008), who achieved high-wavelength precision when using the comb for solar observations. Although there is little doubt that such femtosecond frequency combs offer the highest possible precision, they are not yet easily available and are relatively expensive (though their cost is expected to come down in the future). While facility-class instruments on large telescopes will likely be able to invest in such calibrators, they are not as easy to obtain, maintain and operate as Th-Ar or other emission lamps. Many high resolution NIR spectrographs may also not need this level of sub 1 ms$^{-1}$ wavelength calibration since their performance may be dominated by other systematics. The atmospheric absorption and OH emission lines may also set the limit to the precision that can realistically be achieved in the NIR.
Recent progress in generating THz spacing frequency combs (Del'Haye et al. 2007) by using a micro-toroidal oscillator and a laser may offer some promise in making an inexpensive, readily available device for future NIR spectrographs.
\section{Absorption Gas Cells}
 Absorption gas cells have a long history of usage for calibration. Molecular absorption caused by rotational and vibrational energy levels is well understood, stable, and repeatable. The \citet{CW79} radial velocity survey used the sharp isolated absorption lines of \isotope[]{H}\isotope[]{F} gas in the red to calibrate out instrument drifts. The use of molecular Iodine (I$_2$) gas as a simultaneous absorption calibrator by \citet{Butler96} has lead to the ability to measure very precise velocities, with long term precision approaching 1 ms$^{-1}$, and the stability of such cells is now widely accepted. Absorption cells are not quite as popular as Th-Ar in the UV-optical region because the lines tend to be localized in a smaller wavelength regimes (eg. 500-620nm for I$_2$). In the case of I$_2$, the lines are all blended even at spectral resolutions of 100k, requiring FTS spectra of that particular cell and PSF modeling to determine a wavelength scale when used with a spectrograph. Absorption cells are in use in some astronomical instruments in the infra-red for wavelength calibration. The CRIRES instrument on the VLT uses a \isotope[]{N}$_2$\isotope[]{O} gas cell as a calibrater for wavelengths longer than 2.2 $\mu$m \citep{Kerber08}. Gas mixtures for an I$_2$-like absorbtion gas cell for the planned NAHUAL spectrograph have been discussed by \citet{Martin05} and for the planned GIANO spectrograph by D'Amato et al. (2008) who have developed a combined HCL-HBr-HI cell that has $\sim$200 lines, spanning J, H, \& K bands. \citet{Ramsey08} also mention plans to use \isotope[]{H}\isotope[]{F} and water vapour cells in their J-band pathfinder instrument to monitor velocity drifts. HF has deep lines, but they are few, and the gas is corrosive and difficult to work with. Such a cell also needs to be heated to more than $70\,^{\circ}\mathrm{C}$ to prevent polymerization of HF \citep{CW79}, resulting in line broadening. Water vapour also has many lines, but is also present in the atmosphere, making identification of isolated individual lines for wavelength calibration quite challenging.

 In the NIR H band ($\sim$ 1.45-1.8 $\mu$m) a number of absorption calibrators with well spaced lines have been well characterized,  primarily owing to the needs for ever higher wavelength division multiplexing in the telecommunication industry. Although no single cell covers the entire spectral region, we show that a significant fraction of the H band can be covered using combinations of four commercially available cells. Our proposed use is not to pass the starlight through the cells, but to use the isolated absorption lines as wavelength markers to determine the dispersion solution and track instrument drifts, thereby enabling the measurement of precise velocities and accelerations.

\subsection{NIST SRMS} The United States National Institute of Standards \& Technology (NIST) has designated four standard reference materials (SRMs hereafter) for use in wavelength calibration. Together these four materials span the C \& L telecom wavelength bands and have $\sim 200$ isolated sharp lines in the 1510-1630 nm wavelength region of the H band. Table 1 lists these NIST SRM gas cells, their effective wavelength coverage and the number of lines with NIST-certified wavelengths. These gas cells are commercially available, and have an effective absorption path length of 5-80 cm. The low pressure H\isotope[13]{C}\isotope[14]{N} (Swann \& Gilbert 2000) and \isotope[12]C$_2$H$_2$ \citep{SG05} cells have been primarily designed for high resolution application and have narrow linewidths of 7-15 pm, essentially unresolved at spectral resolutions as high as R=50-70k at 1550nm. The high pressure CO gas cells were originally designed for use with an instrumental resolution of 0.05nm and have linewidths of 50pm \citep{SG02}. Decreasing the cell pressure can easily make the lines sharper, if necessary, and increasing cell length can increase absorption for unsaturated lines. These gas cells are available coupled to single-mode fibers with standardized FC/PC connectors for light input and output, allowing the CO gas cells to be used in a multi-pass configuration, achieving 80 cm of absorption length with four passes through a compact 20 cm cell. The use of the single mode fiber of input and output makes it easy to pass continuum light through multiple cells to obtain an imprint of the absorbtion lines over larger wavelength regimes. The use of the fibers is not necessary, but is a convenience.
\subsection{Astronomical Echelle Spectrographs} Modern high resolution spectrographs generally use coarse ruled echelle gratings that have high blaze angles (R2-R4 or $\tan{\theta_B}=2-4$ is fairly typical). To estimate the achievable accuracy of wavelength calibration, we consider an R2 echelle spectrograph capable of a resolution of R=50k at 1600 nm.
%(similar to one in Ge at al. 2006).
We assume a 4 pixel sampling the FWHM of the Gaussian resolution element. We assume a grating with a groove density such that a wavelength of 1550nm corresponds to an order number of $\sim 100$. While such low groove densities are not available in commercially ruled echelles, they can be created on silicon immersion gratings \citep{Ge06} enabling high dispersion as well as the ability to pack all orders from the H band into a single 1k by 1k detector. Figure 1 shows the expected absorption spectra from these gas cells with such an instrument. For HCN and C$_2$H$_2$ we have used the high resolution spectra scans from \citet{SG00,SG05}\footnote{Also available at www.nist.gov/srm} and convolved them with a Gaussian instrument profile (R=50k at 1600nm). For CO the full spectral scans are not available and we have simulated the data using line centers and broadening coefficients from \citet{SG02}. We have not convolved the simulated CO data with the Gaussian instrument profile since the measurements were obtained at a lower resolution than R=50k. We would expect these lines to be slightly deeper when observed at the resolution we assume for the spectrograph. Figure \ref{fig:gascells} shows a schematic arrangement of the gas cells, coupled with optical fibers, that can be used to generate the absorption spectra seen in Figure 1.

The effective free spectral range of each order, $\sim \lambda/m$, corresponds to ~15nm bandwidth at 1550nm. As can be seen from Figure 1, the HCN cell alone provides 20 sharp reference lines within $\pm8$ nm of the order center, a region where the blaze efficiency is expected to be $> 40$\% of peak \citep{Schroeder}. In reality the echelle order spans a larger wavelength coverage than  $\sim \lambda/m$ on the CCD, making more lines accessible for wavelength calibration. The design of echelles also leads to some wavelength overlap between adjacent orders, allowing the same line to be used to calibrate more than one order. Gratings operating at lower order numbers will have even higher line densities per order if the orders fit on the detector array.
% While the purpose of this letter is to explore the general use of absorption cells, we use a specific example to illustrate the achievable wavelength %accuracy. We model the wave
%length coverage of an instrument using an R2 silicon immersion grating whose performance has been reported in Ge et al. (2006).

  Typical echelle spectra span 7-12 pixels in the slit direction, and we assume 7 pixels here. Assuming that the continuum light source is bright enough (discussed in detail later), a typical continuum S/N of 200 per pixel is a reasonable assumption, assuring also that one is operating in the linear regime in NIR detectors (which typically have quantum wells of ~100k). The data reduction process will optimally combine all 7 pixels in the slit direction into a single pixel with a S/N of $200\sqrt{7}$, or S/N $\sim 500$ per pixel, in the extracted one-dimensional spectrum. Although the actual shape of the absorption line is a Voigt profile, we can approximate the line as a Gaussian for the purpose of estimating the velocity precision. Following the procedure outlined in \citet{Butler96} for calculating the photon-noise limited velocity precision yields an error of ~24 m s$^{-1}$ for a line with a depth (after convolution with the spectrograph instrument profile) of 20\%. So the use of $\sim$20 lines reasonably well spaced across the echelle order can, in principle, enable wavelength calibration accuracy at the level of $\sim 5$ m s$^{-1}$ per echelle order. This uncertainty is smaller than the photon noise error expected per order on most stars observed, and is therefore sufficient. Many HCN and C$_2$H$_2$ lines are deeper than 30\%, while the CO lines have lower depths of $\sim 10-18$\% and have line widths broader than the spectrograph resolution, leading to typical errors of 25-50 m s$^{-1}$ per line. However, these calculations are specifically for the NIST SRM cells and  custom-made longer pathlength CO cells (or more multiple passes through the same cell) can create deeper absorption lines if they are required. We speculate that significantly larger path lengths than provided by conventional cells may be achievable in the future by directly using long lengths of gas-filled hollow-core photonic crystal fibers now under development \citep{Benabid05, Tuominen}.
These gas cells can also be used at higher (or lower) spectral resolutions than the R =50-70k we have considered here. To demonstrate this, we follow the prescription of Bouchy, Pepe \& Queloz (2001) to calculate a Quality factor (Q) that is an estimate of the intrinsic radial velocity information contained in the spectrum. The Q factor depends on factors like the spectral richness of the wavelength region being considered, the sharpness of the absorption lines, as well as the instrument profile. For a given wavelength region, the limiting radial velocity precision ($\sigma_{RMS}$) is given by
\begin{equation}
\sigma_{RMS} = \frac{c}{Q\sqrt{N}},
\end{equation}
where $c$ is the velocity of light, $Q$ the Quality factor, and $N$ the total number of photons collected. As expected, the radial velocity precision is proportional to $\sqrt{N}$.

We calculate the Q factor for the HCN SRM for different spectral resolutions using the wavelength range 1524-1565nm. Figure \ref{fig:qfactor} shows the Quality factor as a function of the spectral resolution. At low spectral resolutions all the individual lines blur into each other, leading to very low velocity information content. At intermediate spectral resolutions (R=50-100k) the information content is a strong function of spectral resolution, as the absorption lines, which have line widths of 7-12pm, are still unresolved. When the spectral resolution exceeds 200,000, the individual lines themselves begin to be resolved and the Q factor begins to plateau, asymptotically approaching its finite value at very high resolutions. Beyond a resolution of R=300k the lines are fully resolved, and the Q factor increase is minimal. Currently planned NIR high-resolution instruments all operate in the R=20-100k regime, and the HCN \& C$_2$H$_2$ SRM cells are well-suited to these. The commercially available CO cells, however, have larger line-widths, making them less desirable for the higher spectral resolutions. Custom cells, however, can easily be manufactured.  Other than high-resolution spectrographs, the absorption cells can be used in other NIR instruments. Planned H band spectrographs like the fiber-fed APOGEE instrument for SDSS-III (Allende-Prieto et al . 2008), or dispersed fixed-delay interferometer instruments (Guo et al. 2006) in the NIR, can use such cells for wavelength calibration.

It is important to recognize that the absorption line spacings in these cells are sparse when compared to absorption line densities in typical M dwarf spectra. The high continuum S/N is absolutely necessary to ensure that the error in the wavelength calibration is smaller than the achievable precision on M dwarfs. This is one of the reasons we have discussed this approach only in the context of a fiber-fed instrument. If such cells were to be used for simultaneous calibration by passing starlight through them, then the achievable precision may be severely limited by the reference rather than the star. Such an approach should be considered only when the need to calibrate out instrument drifts is more important than achieving close to photon noise limited precision on the stellar target.

\subsection{Wavelength Accuracy \& Stability:} The vacuum wavelengths of the line centers for the species in the SRMs can, and have, been measured very accurately at low temperatures and pressures. For example, the measured vacuum wavelengths at low pressure for the $3\nu$ lines of \isotope[12]{C}\isotope[16]{0} \citep{PG97} agree with the theoretically calculated values from the HITRAN database \citep{Rothman05} to 0.02 pm (or $< 4$ m s$^{-1}$ at 1.6 um). For higher cell pressures, the lines begin to broaden and also shift due to interaction of molecules during elastic collisions. Accounting correctly for this pressure shift is the dominant source of uncertainty for determining the line centers of the absorption lines. This shift is accounted for by explicitly measuring it at various pressures and extrapolating a linear relationship. The resulting line centers for the higher pressure CO gas SRMS are certified by NIST to 0.4-0.7 pm (all $2\sigma$ errors), HCN from 0.04 - 0.24 pm and C$_2$H$_2$ from 0.1-0.6 pm \citep{SG00,SG02,SG05}. So the line centers themselves are known to an accuracy of 4-60 m s$^{-1}$ ($1 \sigma$ errors), quite comparable with the Th-Ar emission line errors quoted by PE83. A significant fraction of this error is actually the uncertainty in the pressure of the cells themselves, which may be possible to better constrain with custom-made cells. If a higher accuracy is really necessary, then high-resolution FTS spectra can be used to independently calibrate the cells. Instruments known to possess high stability over a few hours can also take the approach of \citet{LP07} by acquiring many Th-Ar exposures to build up high S/N to determine a wavelength solution, and using that solution to determine the line centers of the absorption cells.

 Of perhaps more concern is the inherent stability of the absorption lines themselves. A change in operating temperature affects the pressure of the cell, which, due to the pressure shift, leads to a shift of the line center. The wavelength shift $\Delta \lambda$ due to pressure at temperatures $T$ and $T_m$ are related by \citep{SG00}
 \begin{equation}
 \Delta \lambda (T) = \Delta \lambda (T_m) \sqrt{(T/T_m)}  .
 \end{equation}
  A one degree change in temperature at a operating temperature of 296 degrees leads to a 0.17\% change in the pressure-induced wavelength shift. For the high pressure CO cells, the pressure-induced shift is at most 3pm, and that one degree change leads to a line shift of $\sim0.005$pm, which is a sub-ms$^{-1}$ effect. Given the fairly loose temperature requirements the entire enclosure containing the cell assembly (Figure \ref{fig:gascells}) could be temperature stabilized to $\pm 1$C using heating elements running in a proportional-integral-derivative (PID) loop with temperature sensors. If necessary, temperature control to $\pm 0.1$ degrees is easily achieved if each cell is stabilized independently with a heating element wrapped around it,and can make temperature effects negligible. The absorption cells are relatively immune to any other environmental effect since they are contained in sealed glass tubes. Unlike Th-Ar lamps, absorption cells do not have a warm up period (typically 15-30 minutes) where the lines are not stable enough for precision applications, or a limit on their operating lifetime. They are also passive, requiring only a light source for operation.
\subsection{Bright White Light Source}
 The single-mode fiber-coupled gas cells makes for a compact and easily configurable set of wavelength references, but the narrow core and spatial filtering properties of such fibers make them inherently difficult to couple to traditional illumination sources like quartz lamps. A number of white light sources coupled to single mode fibers are now commercially available for the telecom S, C and L bands (1400-1630nm), which are well matched to the spectral regime covered by these refereence cells. Alternatively, tunable diode lasers are also available that can be used to scan through the regions of interest at many times per second. Such sources may easily be adaptable for spectroscopic applications in astronomy. The science fiber coupled to the spectrograph needs to be multi-mode to couple effectively to the telescope. To ensure that the science and calibration fiber illuminate the spectrograph optics in the same way, the calibration fiber should be identical to the science fiber. Expanding the light from a single mode fiber to the correct focal ratio and coupling into the multi-mode science and calibration fibers poses no major challenge if this is deemed necessary.
 As mentioned before, the use of single-mode fibers is not mandatory if multi-pass configurations are not required. In such a case, a multi-mode fiber is adequate, still enabling the light to be passed through multiple cells. The penalty of this approach is the unwieldy 80 cm length of the \isotope[]{C}\isotope[]{O} cells, and the advantage is that a conventional quartz-lamp light source is quite sufficient.
 \subsection{Isotopologues}
  The NIST standards are deliberately designed to span as large a range as possible in the telecom bands. For such applications, line densities are not as critical as for the broadband wavelength calibration application in astronomy. Available line densities and wavelength coverage can be increased by using isotopologues of these molecules, which exhibit similar levels of stability but have absorption lines at different wavelengths. For CO the NIST cells are specifically chosen to be \isotope[12]{C}\isotope[16]{O} and \isotope[13]{C}\isotope[16]{O}. The HITRAN database contains line transition parameters for 4 other CO isotopalogues: \isotope[12]{C}\isotope[17]{O}, \isotope[12]{C}\isotope[18]{O}, \isotope[13]{C}\isotope[17]{O}, and \isotope[13]{C}\isotope[18]{O}. Figure \ref{fig:CO} shows the absorption lines for all 6 CO isotopologues listed above. Together, they substantially increase the line densities available for calibration in this region of the H band, and extend coverage to longer wavelengths. The filled circles in the figure correspond to zero-pressure line centers from the HITRAN database. Transmission values are all scaled to that of \isotope[12]{C}\isotope[16]{O} because the purpose of the figure is to illustrate the increase in density and coverage. Actual transmission is best derived from experimental data for such cells and the length required for gas cells of the four additional isotopologues may be different than those of \isotope[12]{C}\isotope[16]{O} and \isotope[13]{C}\isotope[16]{O}. Even higher line densities and coverage than shown here may be possible using \isotope[14]{C}\isotope[]{O}, even though \isotope[14]{C} is unstable and undergoes $\beta$ decay to \isotope[14]{N} with a half-life of $\sim 5730$ years. Similarly the many known isotopalogues of \isotope[12]{C}$_2$H$_2$ and HCN can also be used to generate additional absorption lines. While the use of isotopologues is attractive for increasing line densities, the pressure-shifted line centers for many of these species need to be determined before they can be used. Bootstrapping a wavelength solution off the known NIST calibrated lines is possible for very stable instruments, as described earlier. In general, however, the isotopologue gas cells should also be calibrated independently, like the NIST SRMS.

\subsection{Additional Molecular References in the NIR J \& H bands}
Thus far we have primarily discussed the use of the NIST absorption gas cells and their isotopologues in the NIR H band. As discussed earlier, gas cells calibrations in this wavelength regime exist primarily due to the needs of the telecommunication industry. Although such well-characterized cells are not commercially available in the J band, it is worth briefly discussing the molecules that are promising possibilities in this wavelength regime as well. Hydrogen Fluoride has a series of sharp absorption lines in the 867-909nm and 1257-1340nm regions. Methane and water have lines in the J \& H but both are also atmospheric species. D'Amato et al. (2008) have demonstrated that HCl, HBr and HI exhibit absorption lines in the 1.2-1.4$\mu$m region, though getting deep absorption lines requires path lengths approaching 1 meter. Fiber fed gas cells similar to the NIST SRMs can be easily used to provide the path lengths necessary for moderately deep absorption lines in such cases.
HCL has a series of lines in the 1185-1240nm region and 1720-1870nm regions, which well-complements the molecules we have discussed for the H band, and spans parts of the J band.

\subsection{Safety Considerations in Dealing with Gas Cells}
For practicality, the gas cells must not be toxic or lethal since a leak, or a breakage, would significantly impact operations. Acetylene is not known to be toxic even if inhaled in large concentrations. Hydrogen Cyanide is lethal in large amounts since it inhibits enzymes in the electron transport chain of cells, thereby making normal cell functioning impossible. The trace amounts of HCN found in the NIST SRM cells is quite safe. Even if the HCN contents on the {\it entire} SRM gas cell were to be respired- the increase in the cyanide content in blood would be minimal. Inhaled Carbon Monoxide is lethal in large concentrations since it binds to haemoglobin, preventing oxygen transport. However CO is also harmless in the small trace amounts found in the gas cells. Methane and water are commonly occurring atmospheric species and are not toxic. Hydrogen Fluoride is very corrosive and attacks glass. HF gas cells have to be specially designed with materials that do not react with the gas and the cells have to be handled carefully. Such HF cells have been in use for calibration for a number of years now. The gas cells we have explored here are all safe to use in a laboratory or a confined environment and breakages or leaks, if they happen, are not a cause for major concern.

\section{Discussion}
 We have explored the relative advantages of different wavelength calibration techniques for high-resolution fiber-fed NIR echelle spectrographs. We conclude that a series of commercially available absorption cell standards can be used to wavelength calibrate echelle data over a significant fraction of the H band, covering over 120nm with four gas cells. Some of these cells require long path lengths, but the use of single-mode fibers enables compact multi-pass configurations with small diameter cells that can easily be integrated into a calibration unit. Although the absorption lines are very stable, their absolute line centers are not known a-priori to better than 4-60 m s$^{-1}$. In principle, this uncertainty can be redressed by explicitly acquiring a high-resolution FTS spectrum or measuring them with tunable diode lasers. In practice, even this solution may not be necessary for most applications. We hope to demonstrate the achievable precision of these cells with FIRST, a high-resolution silicon immersion grating based instrument in the preliminary design stage \citep{Ge06}, and to further explore their potential use as calibrators. We have also considered only the four commercially available NIST cells. By using additional gases, isotopologues and cell lengths, one may be able to extend this technique to span larger regions of the H band with more and deeper absorption lines. Many NIR instruments in the design or construction stage may benefit from using the calibration approach we have outlined here. Such an approach may, in concert with Th-Ar lamp used simultaneously for regions without a absorption reference, be able to provide adequate wavelength calibration for most high precision applications until frequency-comb technology is mature, commonly available, and less expensive.

We are very grateful to Dimitri Veras, Curtis DeWitt and Fred Hearty for useful discussions and a careful reading of this manuscript. We thank Steve Blazo, Wavelength References, for useful discussions about multi-gas cells and light sources. We acknowledge the support from NSF with grant NSF AST-0705139, NASA with grants NNX07AP14G and NNG05GR41G, UCF-UF SRI program, and the University of Florida.

\begin{table}
\caption{Commercially Available NIST Wavelength Reference Gas Cells suitable for H band wavelength calibration.}
\begin{tabular}{ l l c c c}
\hline
NIST SRM & Species & Range (nm) & Pressure (Torr) & Lines \\
\hline
2517a & \isotope[12]{C}$_2$H$_2$ & 1510-1540 & 50  & 56 \\
2519a & H\isotope[13]{C}\isotope[14]{N} & 1530-1560 & 25 & 51 \\
2514 & \isotope[12]{C}\isotope[16]{O}  & 1560-1595 & 1000 & 41 \\
2515 & \isotope[13]{C}\isotope[16]{O}  & 1595-1630 & 1000 & 41 \\
\hline
\end{tabular}
\end{table}

\begin{figure}%[htbp]
\begin{center}
\includegraphics*[width=5.2in,angle=0]{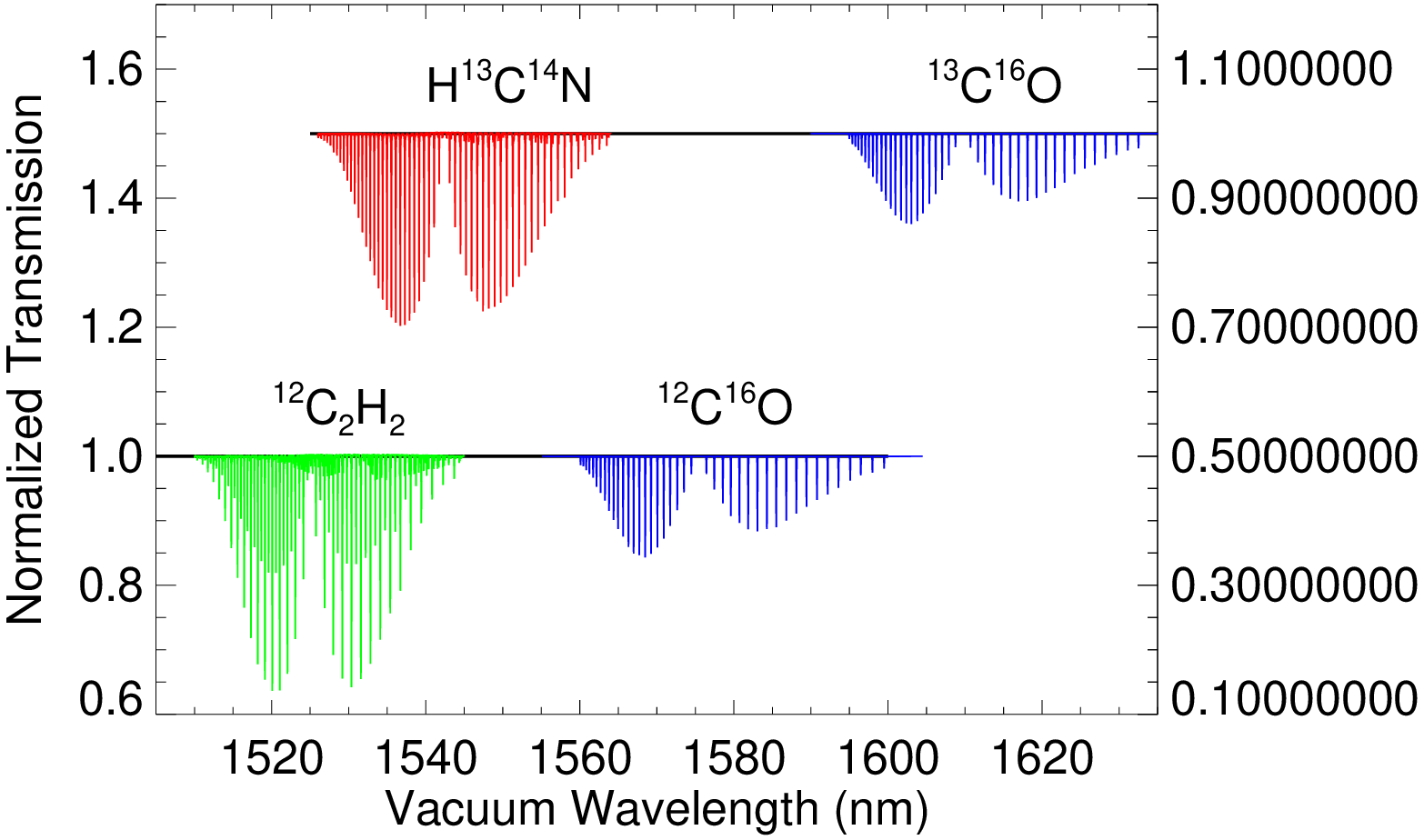}
\end{center}
\caption[NIST SRM for Acetylene]{Expected transmission for the NIST SRM cells when observed with a spectrograph with a spectral resolution of 50k. For HCN and C$_2$H$_2$, we have simulated the transmission at this resolution using available high resolution data of the NIST SRMs. The spectra shown are for a 5 cm long C$_2$H$_2$ cell and for a 15 cm long HCN cell.  For CO, the spectrum is simulated to resemble the spectrum observed by Swann \& Gilbert (2002) using a 0.05nm bandwidth spectrum analyzer (effectively R=32k). The CO lines are expected to be a few percent deeper than shown here when observed at R=50k. While the length of the CO cells is 20 cm, the light makes four passes through the cell for a total absorption path of 80 cm.} \label{fig:kpnodrift}
\end{figure}

\begin{figure}%[htbp]
\begin{center}
\includegraphics*[width=5.2in,angle=0]{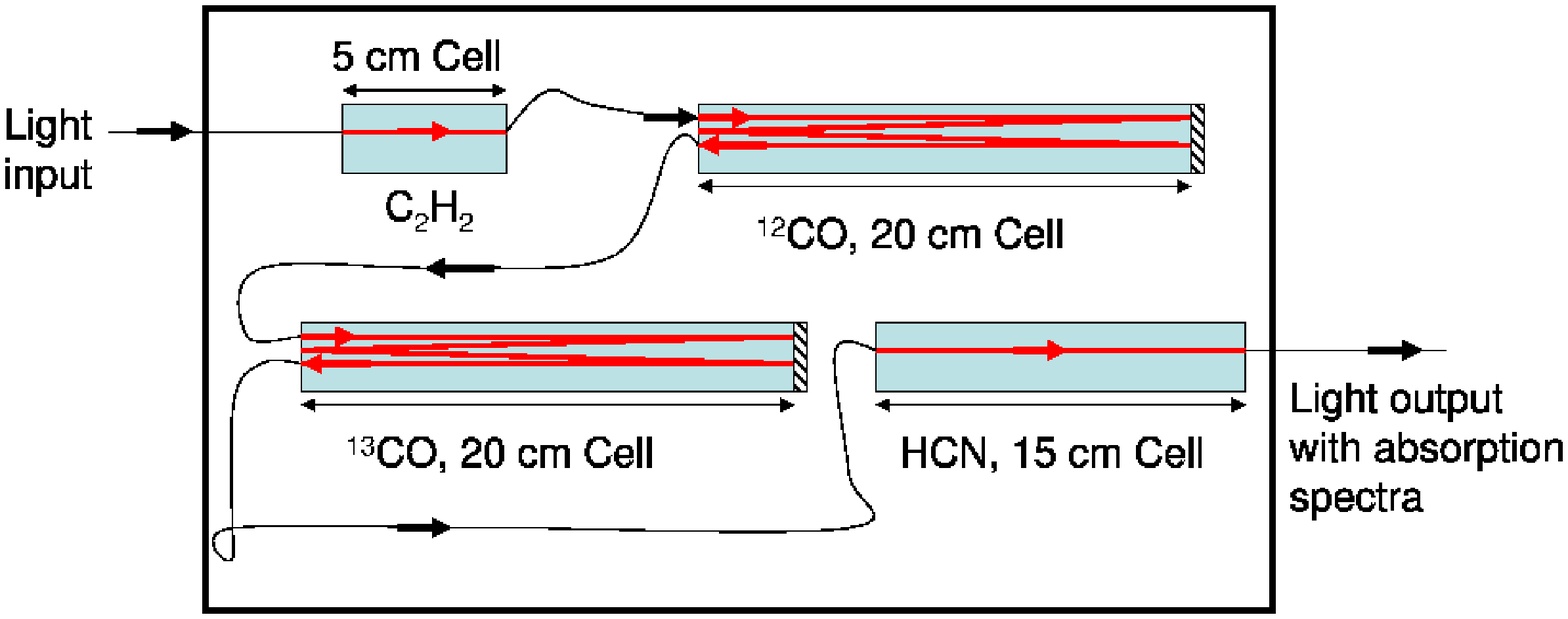}
\end{center}
\caption[]{A schematic showing the SRM gas cells, their lengths and an arrangement of cells that yields a transmission spectrum similar to that in Figure \ref{fig:kpnodrift}. The schematic illustrates the single-pass transmission for HCN and C$_2$H$_2$ gas cells as well as the 4-pass transmission in the 20 cm CO gas cell to yield a 80 cm absorption path length. All cells are linked with optical fibers. For clarity we have shown the multi-pass arrangement as four separate beams, while in reality two sets of the beams are parallel. A full diagram of the CO SRM cell and holder can be found in Figure 2 of NIST publication 260-146 \footnote{http://ts.nist.gov/MeasurementServices/ReferenceMaterials/upload/SP260-146.PDF }} \label{fig:gascells}
\end{figure}

\begin{figure}%[htbp]
\begin{center}
\includegraphics*[width=5.2in,angle=0]{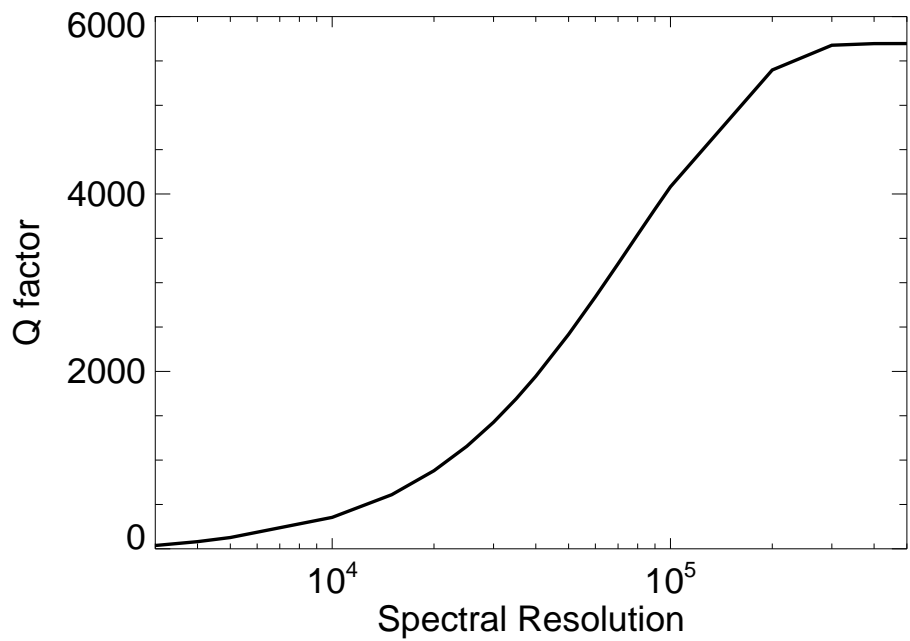}
\end{center}
\caption[Q factor for HCN]{The Quality Factor (Q) for the HCN SRM as a function of the spectral resolution. The quality factor is an estimate of the radial velocity related information content in the spectra. The increase in the Q factor begins to plateau as the width of the resolution element becomes significantly smaller than the intrinsic line-widths in the spectra.} \label{fig:qfactor}
\end{figure}

\begin{figure}%[htbp]
\begin{center}
\includegraphics*[width=5.2in,angle=0]{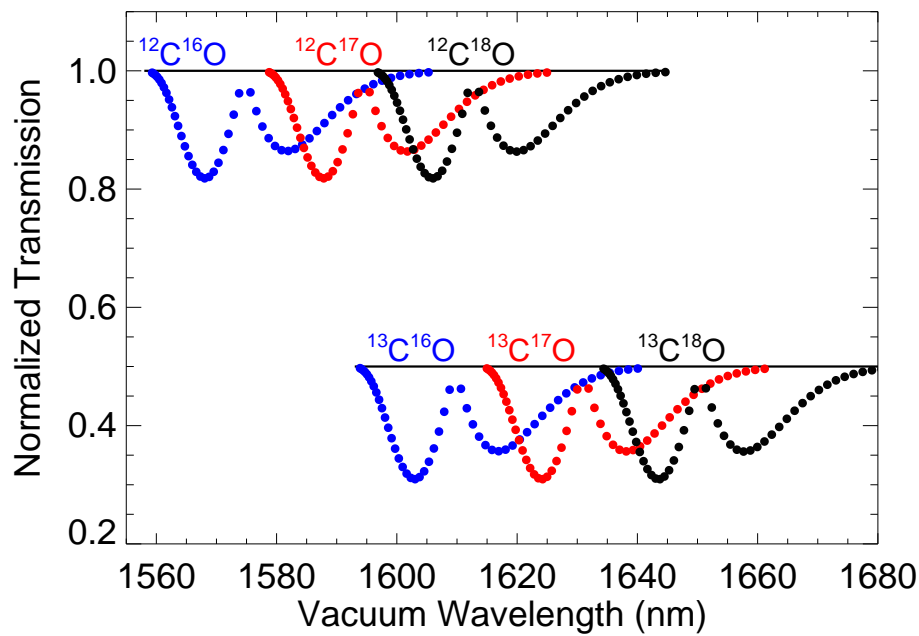}
\end{center}
\caption[Isotopalogues]{Absorption spectra of isotopologues of \isotope[]{C}\isotope[]{O} that are present in the HITRAN database. Points correspond to the zero-pressure line centers listed in the database. Transmission spectra of all isotopologues has been scaled to that of \isotope[12]{C}\isotope[16]{O} for ease of comparison. The transmission values on the y-axis are for illustration only.} \label{fig:CO}
\end{figure}

\begin{thebibliography}{}
\bibitem[Allende Prieto et al.(2008)]{2008arXiv0809.2362A} Allende Prieto,
C., et al.\ 2008, ArXiv e-prints, 809, arXiv:0809.2362
\bibitem[Baranne et al.(1996)]{Baranne96} Baranne, A., et al.\ 1996, AAPS, 119, 373
\bibitem[Benabid et al.(2005)]{Benabid05} Benabid, F., Couny, F.,
Knight, J.~C., Birks, T.~A., \& Russell, P.~S.~J.\ 2005, Nature, 434, 488
\bibitem[Bouchy et
al.(2001)]{2001A&A...374..733B} Bouchy, F., Pepe, F., \& Queloz, D.\ 2001, \aap, 374, 733
\bibitem[Braje et al.(2008)]{Braje08} Braje, D.~A., Kirchner,
M.~S., Osterman, S., Fortier, T.,
\& Diddams, S.~A.\ 2008, European Physical Journal D, 48, 57
\bibitem[Butler et al.(1996)]{Butler96} Butler, R.~P., Marcy,
G.~W., Williams, E., McCarthy, C., Dosanjh, P., \& Vogt, S.~S.\
1996, PASP, 108, 500
\bibitem[Campbell \& Walker(1979)]{CW79} Campbell, B., \& Walker, G.~A.~H.\ 1979, PASP, 91, 540
\bibitem[Endl et al.(2008)]{2008ApJ...673.1165E} Endl, M., Cochran, W.~D.,
Wittenmyer, R.~A., \& Boss, A.~P.\ 2008, \apj, 673, 1165
\bibitem[Forveille et al.(2008)]{2008arXiv0809.0750F} Forveille, T., et
al.\ 2008, ArXiv e-prints, 809, arXiv:0809.0750
\bibitem[Ge et al.(2006)]{Ge06} Ge, J., McDavitt, D., Zhao,
B., Mahadevan, S., DeWitt, C., \& Seager, S.\ 2006, Proceedings of SPIE, 6269,
\bibitem[Guo et al.(2006)]{2006AAS...209.8513G} Guo, P., Ge, J., Mahadevan,
S.,
\& Ramsey, L.\ 2006, Bulletin of the American Astronomical Society, 38, 1016
\bibitem[Hinkle et al.(2001)]{Hinkle01} Hinkle, K.~H., Joyce,
R.~R., Hedden, A., Wallace, L., \& Engleman, R.~J.\ 2001, PASP, 113, 548
\bibitem[Huelamo et al.(2008)]{2008arXiv0808.2386H} Huelamo, N., et al.\
2008, ArXiv e-prints, 808, arXiv:0808.2386
\bibitem[Hunter
\& Ramsey(1992)]{HR92} Hunter, T.~R., \& Ramsey, L.~W.\ 1992, PASP, 104, 1244
\bibitem[Kerber et al.(2007)]{Kerber07} Kerber, F., Nave, G.,
Sansonetti, C.~J., Bristow, P.,
\& Rosa, M.~R.\ 2007, The Future of Photometric, Spectrophotometric and Polarimetric Standardization, 364, 461
\bibitem[Jaffe et al.(2006)]{2006SPIE.6269E.143J} Jaffe, D.~T., Mar, D.~J.,
Warren, D., \& Segura, P.~R.\ 2006, \procspie, 6269,
\bibitem[Kerber et al.(2008)]{Kerber08} Kerber, F., Nave, G.,
Sansonetti, C.~J., Curto, G.~L., Bristow, P.,
\& Rosa, M.~R.\ 2008, Precision Spectroscopy in Astrophysics, 119
\bibitem[Li et al.(2008)]{Li08} Li, C.-H., et al.\ 2008, Nature, 452, 610
\bibitem[Lovis
\& Pepe(2007)]{LP07} Lovis, C., \& Pepe, F.\ 2007, A\&A, 468, 1115
\bibitem[Mart{\'{\i}}n et al.(2005)]{Martin05} Mart{\'{\i}}n,
E.~L., Guenther, E., Barrado y Navascu{\'e}s, D., Esparza, P., Manescau,
A., \& Laux, U.\ 2005, Astronomische Nachrichten, 326, 1015
\bibitem[Mayor et al.(2008)]{Mayor08} Mayor, M., et al.\ 2008,
ArXiv e-prints, 806, arXiv:0806.4587
\bibitem[Murphy et al.(2007)]{Murphy07} Murphy, M.~T., et al.\
2007, MNRAS, 380, 839
\bibitem[Norl{\'e}n(1973)]{Norlen73} Norl{\'e}n, G.\ 1973, Physica Scripta, 8, 249
\bibitem[Palmer \& Engleman(1983)]{PE83} Palmer, B.~A., \& Engleman, R.\ 1983, LA, Los Alamos: National Laboratory, |c1983,
\bibitem[Pepe et al.(2004)]{Pepe04} Pepe, F., Mayor, M.,Queloz, D., \& Udry, S.\ 2004, Planetary Systems in the Universe, 202, 103
\bibitem[Picqu{\'e}\& Guelachvili(1997)]{PG97} Picqu{\'e}, N., \& Guelachvili, G.\ 1997, Journal of Molecular Spectroscopy, 185, 244
\bibitem[Prato et al.(2008)]{2008arXiv0809.3599P} Prato, L., Huerta, M.,
Johns-Krull, C.~M., Mahmud, N., Jaffe, D.~T.,
\& Hartigan, P.\ 2008, ArXiv e-prints, 809, arXiv:0809.3599
\bibitem[Ramsey et al.(2008)]{Ramsey08} Ramsey, L.~W., Barnes,
J., Redman, S.~L., Jones, H.~R.~A., Wolszczan, A., Bongiorno, S., Engel,
L., \& Jenkins, J.\ 2008, \pasp, 120, 887
\bibitem[Rucinski et al.(2008)]{2008arXiv0809.3987R} Rucinski, S.~M., et
al.\ 2008, ArXiv e-prints, 809, arXiv:0809.3987
\bibitem[Rothman et al.(2005)]{Rothman05} Rothman, L.~S., et al.\
2005, Journal of Quantitative Spectroscopy and Radiative Transfer, 96, 139
\bibitem[Rayner
\& PRVS Team(2007)]{Rayner07} Rayner, J., \& PRVS Team 2007, American Astronomical Society Meeting Abstracts, 211, \#134.19
\bibitem[Scalo et al.(2007)]{Scalo07} Scalo, J., et al.\ 2007,
Astrobiology, 7, 85
\bibitem[Schroeder(1987)]{Schroeder} Schroeder, D.~J.\ 1987, San Diego: Academic Press, 1987
\bibitem[Seifahrt
\& Kaeufl(2008)]{2008arXiv0809.1435S} Seifahrt, A., \& Kaeufl, H.-U.\ 2008, ArXiv e-prints, 809, arXiv:0809.1435
\bibitem[Setiawan et al.(2008)]{2008Natur.451...38S} Setiawan, J., Henning,
T., Launhardt, R., M{\"u}ller, A., Weise, P., Kurster, M.\ 2008, \nat, 451, 38
\bibitem[Steinmetz et al.(2008)]{2008arXiv0809.1663S} Steinmetz, T., et al.\ 2008, ArXiv e-prints, 809, arXiv:0809.1663
\bibitem[Swann \& Gilbert(2000)]{SG00} Swann, W.~C., \& Gilbert, S.~L.\ 2000, Journal of the Optical Society of America B Optical Physics, 17, 1263
\bibitem[Swann \& Gilbert(2002)]{SG02} Swann, W.~C., \& Gilbert, S.~L.\ 2002, Journal of the Optical Society of America B Optical Physics, 19, 2461
\bibitem[Swann \& Gilbert(2005)]{SG05} Swann, W.~C., \& Gilbert, S.~L.\ 2005, Journal of the Optical Society of America B Optical Physics, 22, 1749
\bibitem[Tarter et al.(2007)]{Tarter07} Tarter, J.~C., et al.\
2007, Astrobiology, 7, 30
\bibitem[Tuominen et al.(2005)]{Tuominen} Tuominen, J., Ritari,
T., Ludvigsen, H.,
\& Petersen, J.~C.\ 2005, Optics Communications, 255, 272
\bibitem[Whaling et al.(1995)]{Whaling95} Whaling, W., Anderson,W.~H.~C., Carle, M.~T., Brault, J.~W., \& Zarem, H.~A.\ 1995, Journal of Quantitative Spectroscopy and Radiative Transfer, 53, 1
\end{thebibliography}
\end{document}